# Resolving the structure of TiBe$_{12}$

M. L. Jackson[a,b], P. A. Burr[c], R. W. Grimes[a]

[a] Centre for Nuclear Engineering, Department of Materials, Imperial College London, SW7 2AZ, UK.
[b] Culham Centre for Fusion Energy, Culham Science Centre, Abingdon, Oxfordshire, OX14 3DB, UK
[c] School of EE&T, University of New South Wales, Sydney NSW 2052, Australia

**Abstract**

There has been considerable controversy regarding the structure of TiBe$_{12}$, which is variously reported as hexagonal and tetragonal. Lattice dynamics simulations based on density functional theory show the tetragonal phase space group *I4/mmm* to be more stable over all temperatures, while the hexagonal phase exhibits an imaginary phonon mode, which, if followed, would lead to the cell adopting the tetragonal structure. We then report the predicted ground state elastic constants and temperature dependence of the bulk modulus and thermal expansion for the tetragonal phase.

**Introduction**

TiBe$_{12}$ is a promising candidate for neutron multiplier applications in future fusion reactors due to its exceptional combination of material properties [1]. Based on earlier investigation [2–5], several modern papers equally assume both hexagonal [1,6,7] and tetragonal [8–11] structures. In particular, most of the modelling efforts have been carried out on the hexagonal structure [6,7]. Both structures can be found in the crystallographic structure databases suggesting that they provide equivalent description of the unit cell. However, simple nearest neighbor analysis shows that in fact they are distinct. Which structure corresponds to the reality is not a priory clear and requires special study. Here we review the discussion on the structure of TiBe$_{12}$ and, with the aid of density functional theory (DFT) and quasi-harmonic phonon calculations, investigate the relative stability of both structures.

The crystal structure of TiBe$_{12}$ was first identified by Raeuchle & Rundle [2], from single crystal measurements, as disordered hexagonal with lattice parameters *a* = 29.44 Å and *c* = 7.33 Å. In a later report concerned with the (tetragonal) structure of MoBe$_{12}$, Raeuchle and von Batchelder [3] state that structure of TiBe$_{12}$ was complex and "in fact, is not yet completely known". In the article they further discuss how their investigation into MoBe$_{12}$ provided clues that could refine the TiBe$_{12}$ structure previously reported, and "that the refinement will increase the similarity between the two structures". Subsequent publications by Zalkin *et al.* [4] and then by Gillam *et al.* [5], who used powder samples, reported a tetragonal structure analogous to MoBe$_{12}$ (prototype Mn$_{12}$Th) with lattice parameters $a$ = 7.35 Å and $c$ = 4.19 Å. This structure is also common among iso-stoichiometric transition metal beryllides (e.g. VBe$_{12}$, CrBe$_{12}$, MoBe$_{12}$, WBe$_{12}$) [12]. Furthermore, Gillam *et al.* [10] provided the structural relationship between TiBe$_{12}$ and Ti$_2$Be$_{17}$, which led to their conclusion that "Raeuchle and Rundle's [hexagonal] structure determination was carried out on crystals of Ti$_2$Be$_{17}$ instead of crystals of TiBe$_{12}$" [5]. In fact, the two TiBe$_{12}$ phases are also closely related (figure 1), and produce similar diffraction patterns as demonstrated in figure 2.

Several recent studies [6,7] have used/assumed a sub-cell of the hexagonal crystal structure identified by Raeuchle and Rundle [8] with lattice parameters of $a$ = 4.26 Å and $c$ = 7.33 Å. This differs from the structure defined by the full unit cell in that Ti atoms between alternate sub-cells should be displaced by ±½[0001] in a disordered fashion. As such, this sub-cell is not a true representation of the crystal structure reported by Raeuchle and Rundle [8].

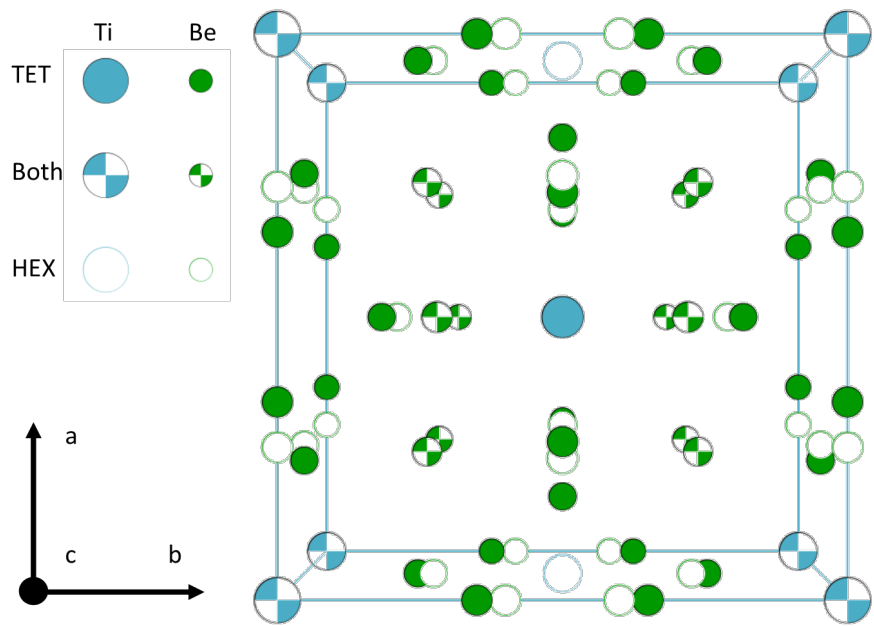

Figure 1. Correspondence between the *I4/mmm* tetragonal structure and the *P6/mmm* sub-cell for TiBe$_{12}$. In *I4/mmm* the 'a' direction corresponds to the 'c' direction in *P6/mmm*. Alternate Ti atoms are displaced by ½ in the *I4/mmm* [100] direction/*P6/mmm* [0001] direction. Be positions are only slightly perturbed.

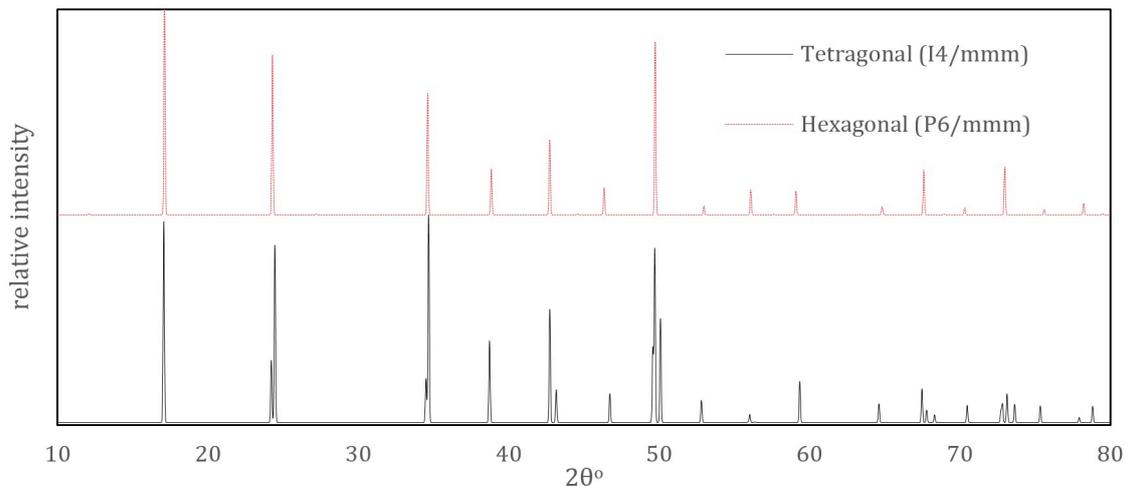

Figure 2. Simulated diffraction patterns for the fully ordered *P6/mmm* and *I4/mmm* unit cells.

## Density Functional Theory Simulations

Both the tetragonal structure and the hexagonal sub-cell were modelled with the Castep [13] DFT code using plane waves with an energy cut-off of 480 eV and ultrasoft pseudopotentials; k-points spacing was kept below 0.3 nm$^{-1}$. The structures were relaxed until atomic forces and stresses were less than $10^{-3}$ eV/Å and $10^{-2}$ eV/Å² respectively.

First the classical ground state of the systems was calculated following eq. 1

$$E_f(\text{TiBe}_{12}) = E^{\text{DFT}}(\text{TiBe}_{12}) - E^{\text{DFT}}(\text{Ti}) - 12E^{\text{DFT}}(\text{Be}) \qquad \text{Eq. 1}$$

The formation energies of the hexagonal and tetragonal phases were found to be −6.82 eV and −7.90 eV per formula unit respectively. A difference in formation energy as large as 1.08 eV suggests that the tetragonal phase is stable at low temperatures. However, zero point energy and temperature effects also contribute to relative stability; these are considered next.

The phonon dispersion curves and density of states (DOS) were computed using the supercell method to evaluate the dynamical matrix from the force constant matrix [14]. Supercells containing 54 and 234 atoms were used for the tetragonal structure and supercells containing 54 and 312 atoms were used for the hexagonal sub-cell. The resulting dispersion curves from the larger supercells are presented in figures 3 and 4. It is evident that the hexagonal structure contains a soft mode at the M q-point – a clear indication of instability in the ground state. Following this mode would lead to a hexagonal/tetragonal transformation as it displaces Ti and a Be in ⟨0001⟩ directions to their corresponding positions in the tetragonal phase.

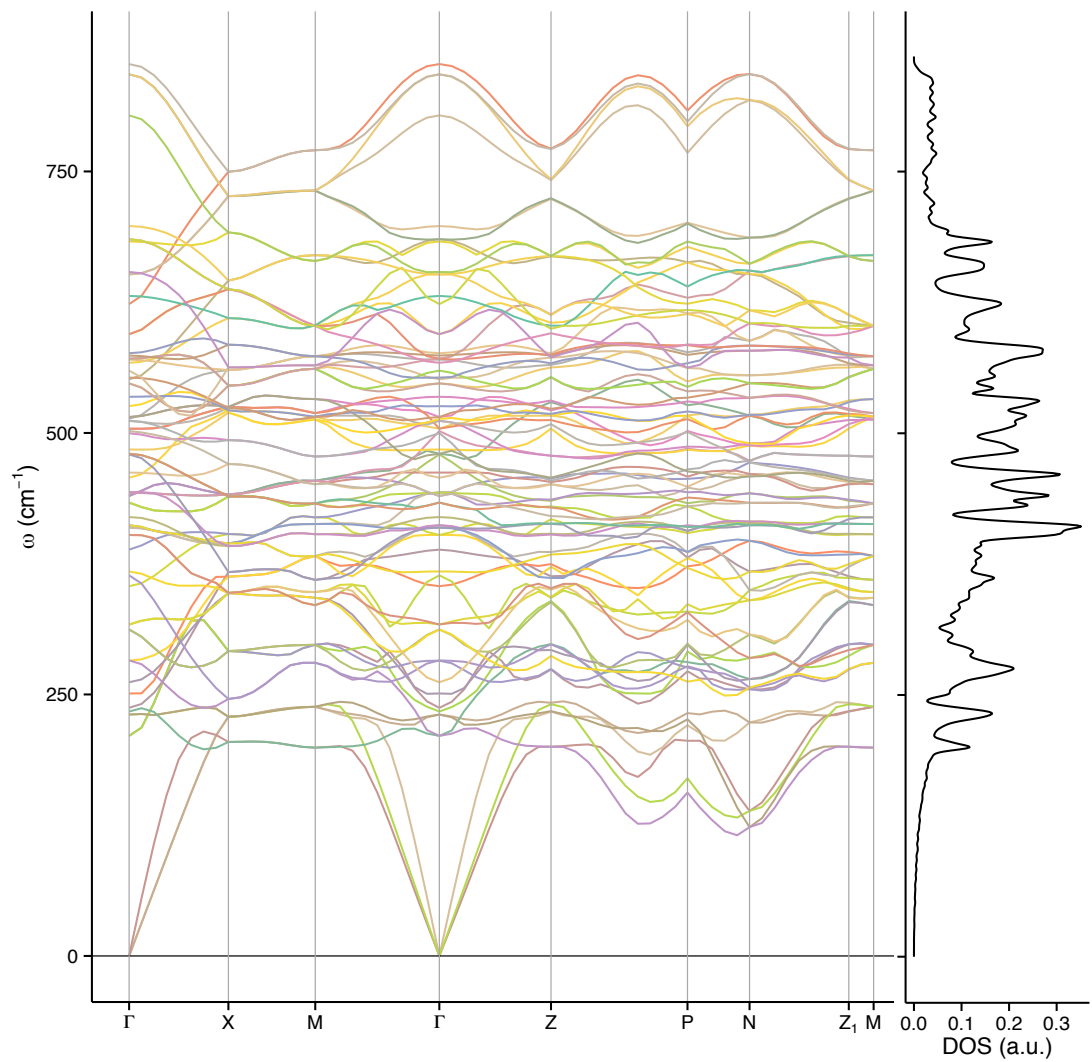

Figure 3. Dispersion curve and DOS of the tetragonal *I4/mmm* TiBe$_{12}$ structure.

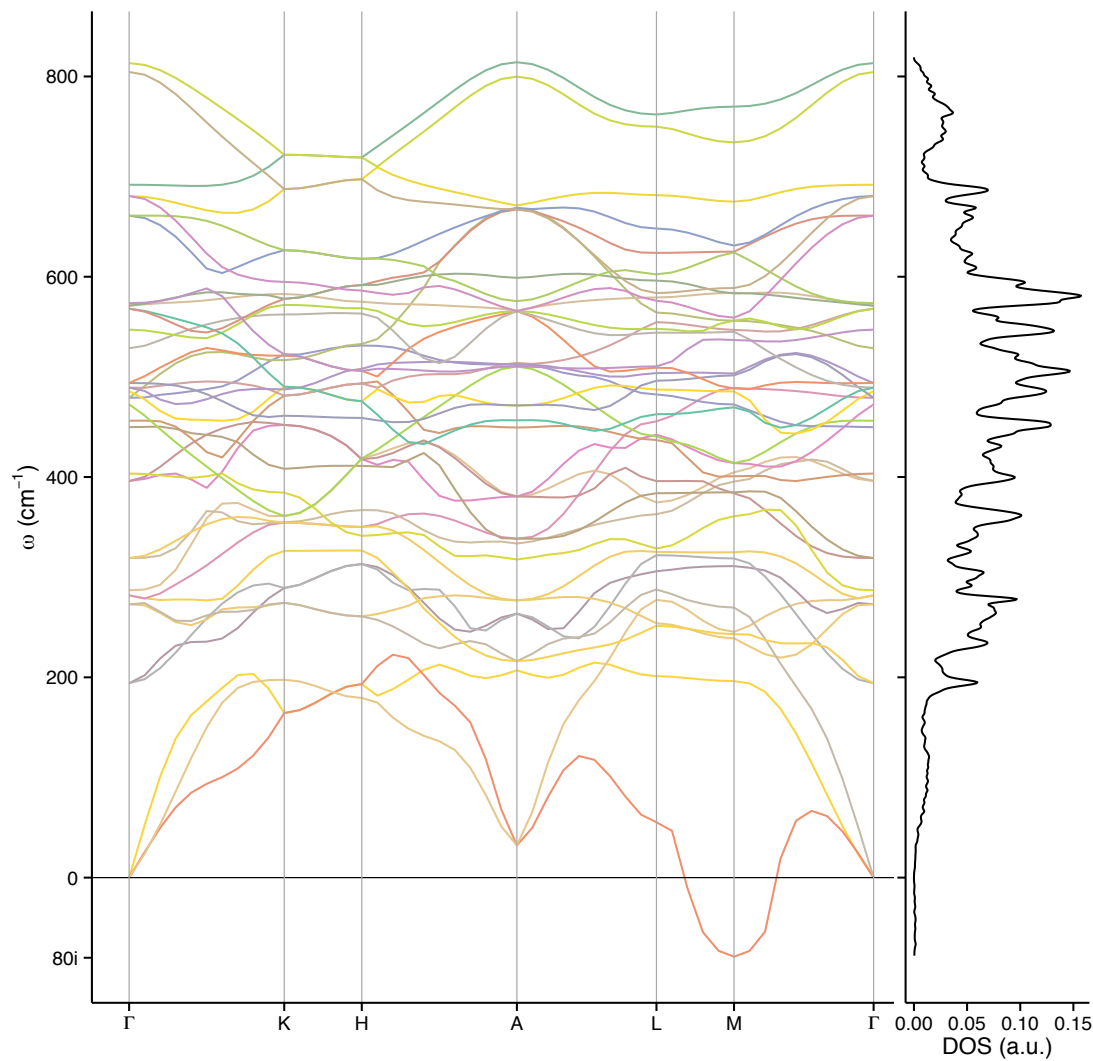

Figure 4 – Dispersion curve and DOS for the hexagonal *P6/mmm* sub-cell of Ti$_{12}$Be.

Zero-point energy and temperature contributions to the Helmholtz free energy were obtained by integrating the phonon DOS, following the harmonic and quasi-harmonic approximations, as outlined in [15]. Figure 5 shows the combined internal (U) and Helmholtz (F) free energy for both systems as a function of temperature up to 1800 K. The excellent agreement between the small supercell and large supercell calculations provide confidence that the 54-atom supercells are adequate for phonon calculations in these systems. Consequently, the quasi-harmonic method, was carried out with the 54-atom supercells only. It is evident that the tetragonal *I4/mmm* structure is consistently more stable compared to hexagonal *P6/mmm*, in line with the thermodynamic instability of the hexagonal structure.

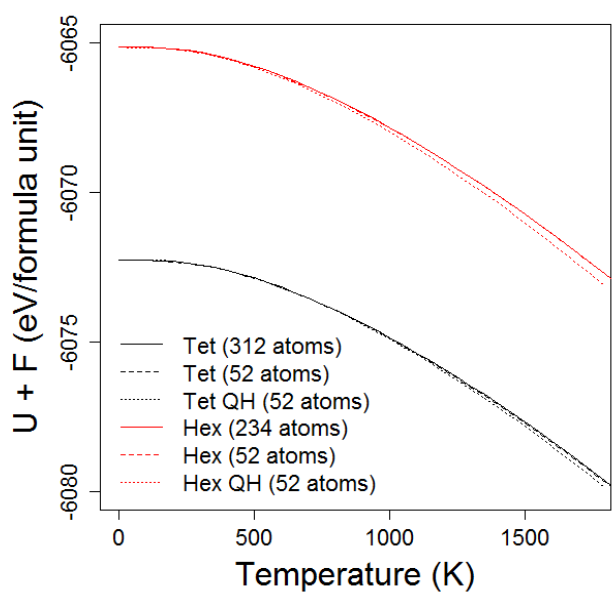

Figure 5. Internal + free energy as a function of temperature. QH=quasi-harmonic. Harmonic results with different supercell sizes appear so close as to be indistinguishable.

**Properties of $I_4/mmm$ TiBe$_{12}$**

Now that the structure of TiBe$_{12}$ is established, we report some of its fundamental materials properties that arise from the quasi-harmonic approach (see supplementary materials for calculation details). These are presented in Figure 6 and Table 1, together with experimental data where available.

The thermal expansion of the tetragonal phase compares favourably to the available experimental data. The bulk modulus follows a typical relationship with temperature for a metal, decreasing with increasing temperature. Predicted lattice constants are slightly overestimated when thermal effects are taken into account, however the bulk modulus is extremely close to the experimental value. This is the first report of stiffness constants for $I_4/mmm$ TiBe$_{12}$.

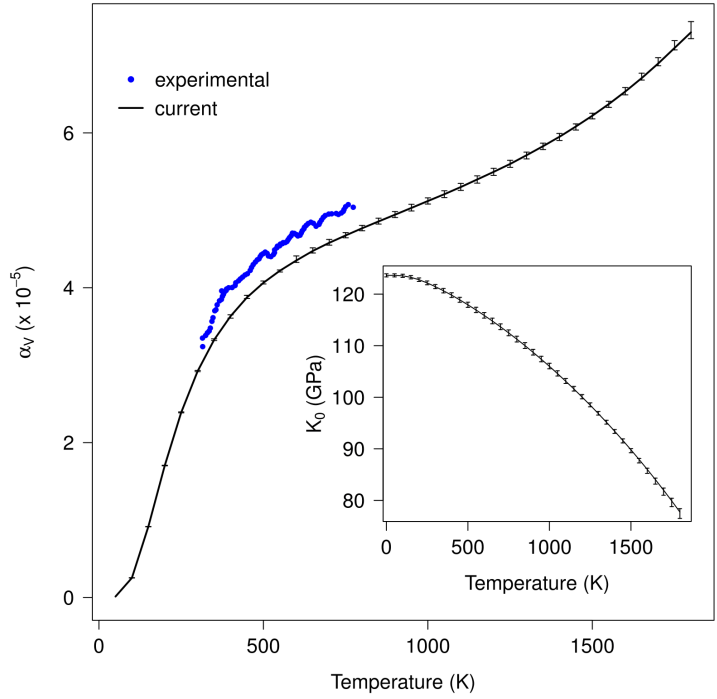

Figure 6. Volumetric thermal expansion coefficient ($\alpha_V$) and bulk modulus ($K_0$) of tetragonal $TiBe_{12}$ [10].

Table 1. Lattice parameters and elastic constants of tetragonal $TiBe_{12}$. Hill's average [16] was used to calculate ground state shear (G) and bulk (K) moduli.

| | a (Å) | c (Å) | $c_{11}$ (GPa) | $c_{12}$ (GPa) | $c_{13}$ (GPa) | $c_{33}$ (GPa) | $c_{44}$ (GPa) | $c_{66}$ (GPa) | K (Gpa) | G (Gpa) |
|---|---|---|---|---|---|---|---|---|---|---|
| DFT ground state | 7.359 | 4.164 | 362.4 | 2.4 | 20.8 | 327.5 | 129.6 | 117.2 | 126.7 | 140.8 |
| DFT QH (T=0K) | 7.446 | 4.216 | - | - | - | - | - | - | 123.6 | - |
| DFT QH (T=300K) | 7.457 | 4.223 | - | - | - | - | - | - | 121.4 | - |
| experimental (T=273K) | 7.35[a] | 4.19[a] | - | - | - | - | - | - | 117.0[b] | - |

a [4] b [17]

**Conclusion**

Using atomic scale quantum mechanical simulations we have investigated the controversy regarding the crystal structure of $TiBe_{12}$, originating from papers published between 1952 and 1964. Lattice dynamics simulations for $TiBe_{12}$ are consistent with the tetragonal structure proposed by Zalkin *et al.* [4] and by Gillam *et al.* [5] (space group *I4/mmm*), not the hexagonal structure proposed by Raeuchle and Rundle [2] (space group *P6/mmm*), or the derivative hexagonal sub-cell that has been used recently in modelling studies [6,7]. While larger supercells are investigated, the 54 atom cell was sufficient for calculating the phonon density of states. Further, for this system little difference was found between harmonic and quasi-harmonic based contributions to the Helmzoltz free energy. Elastic and thermal expansion data for the tetragonal phase are also reported, which are useful for further consideration of the material as a structural component in fusion reactor applications.

## Acknowledgments

M L J thanks CCFE and P A B thanks the EPSRC and ANSTO for financial support. The computing resources were provided by Imperial College HPC.

# Resolving the structure of TiBe$_{12}$ – Supplementary Material

M. L. Jackson[a,b], P. A. Burr[c], R. W. Grimes[a]

[a] Centre for Nuclear Engineering, Department of Materials, Imperial College London, SW7 2AZ, UK.
[b] Culham Centre for Fusion Energy, Culham Science Centre, Abingdon, Oxfordshire, OX14 3DB, UK
[c] School of EE&T, University of New South Wales, Sydney NSW 2052, Australia

Quasi-harmonic thermodynamic data was obtained by repeating the phonon DOS simulation with different unit-cell volumes. The resulting U+F curves (Figure S1) were fitted with a Birch-Murnaghan equation of state [1,2] (Eq. S1).

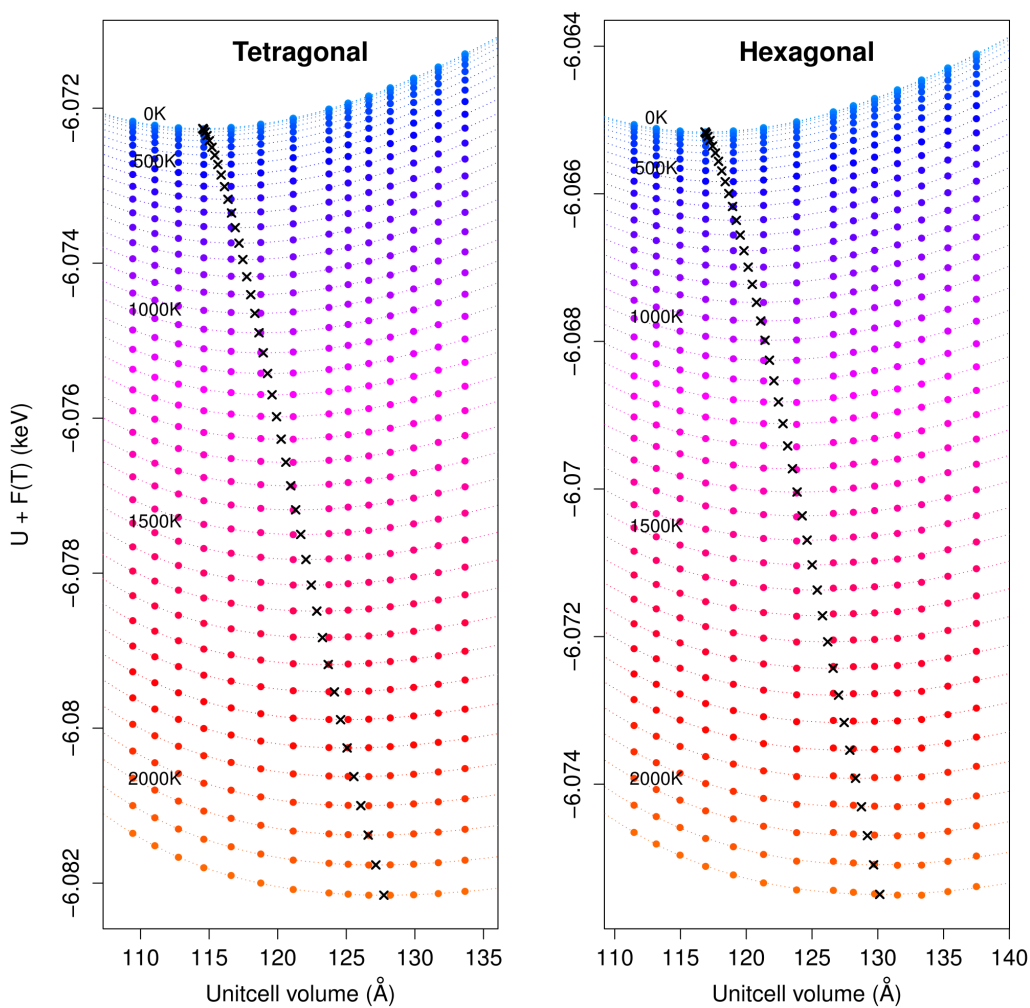

Figure S1 - Thermodynamic data from quasi-harmonic calculations at 50K intervals. Dotted lines are fitted Birch-Murnaghan equations of state, and the crosses represent the minima of those curves.

The Birch-Murnaghan equation (Eq. S1) is represented in the energy-volume form, where $E_0$ is the ground state energy, $V_0$ the reference volume, $V$ the deformed volume, $K_0$ the bulk modulus and $K_0'$ the derivative of the bulk modulus with respect to pressure. From the minima of the Birch-Murnaghan fits, the volumetric thermal expansion ($\alpha_v$) was obtained following equation S2.

$$E(V) = E_0 + \frac{9V_0 K_0}{16} \left\{ \left[ \left(\frac{V_0}{V}\right)^{\frac{2}{3}} - 1 \right]^3 K_0' + \left[ \left(\frac{V_0}{V}\right)^{\frac{2}{3}} - 1 \right]^2 \left[ 6 - 4 \left(\frac{V_0}{V}\right)^{\frac{2}{3}} \right] \right\} \quad \text{Eq. S1}$$

$$\alpha_V = \frac{1}{V} \frac{dV}{dT} \quad \text{Eq. S2}$$